\documentstyle[editedvolume,namedreferences,psfig]{crckapb}  
 
\begin{opening} 
\title{Constraints on particle acceleration from the extragalactic
$\gamma$-ray background} 
\subtitle{} 
 
\author{K. MANNHEIM} 
\institute{Universit\"ats-Sternwarte\\ 
           Geismarlandstra{\ss}e 11, D-37083 G\"ottingen} 
\end{opening} 
  
\begin{document} 
 \begin{abstract}
The propagation of $\gamma$-rays through metagalactic space
is associated with pair creation and subsequent inverse-Compton
scattering off low-energy
background radiation.  As a consequence, 
$\gamma$-rays of very high energy emitted by remote sources
are reprocessed into the window from 10~MeV to 30~GeV 
conserving the injected energy.
Any cosmologically distributed
population of $\gamma$-ray sources therefore contributes to
the  diffuse $\gamma$-ray background in this energy band
which is well-determined from recent observations with the Compton
Gamma Ray Observatory (CGRO).
Since the $\gamma$-rays trace accelerated particles, the observed
flux of diffuse $\gamma$-rays also constrains the global
efficiency for particle acceleration.  
Radio galaxies can account for the $\gamma$-ray background
if their particle acceleration efficiency considerably exceeds $\sim 18$\%
implying that particle acceleration is an essential part
of the thermodynamics in these sources.

\end{abstract}
 
\section{Introduction} 
 
The microphysics of particle acceleration in the solar system is a fascinating
realm and provides useful paradigms for particle acceleration in other places
of the Universe.  From the observed flux of cosmic rays it is clear that much
more powerful particle accelerators than main sequence stars and their winds
must be at work in our Galaxy and in other galaxies.  Supernova remnants 
\cite{ellison97} and
radio galaxies \cite{biermann95} 
are among the most interesting putative sources of cosmic rays.
The cosmic ray pressure is rather large and compares with that of the
interstellar magnetic fields implying that cosmic rays are an important
ingredient in the dynamics of the interstellar medium, and possibly also in the
dynamics of the intergalactic medium (in superclusters).  This imposes strong
constraints on the energetics of possible cosmic ray sources from which one can
infer the particle acceleration efficiency.  For instance, supernova
blast waves can supply the flux of cosmic rays up to the so-called knee in
their spectrum at an energy of $10^{15}$~eV only if their particle acceleration
efficiency is as large as $\sim 13\%$ 
\cite{drury90} which basically invalidates simple
test-particle approaches for the description of the acceleration mechanism.
This has enforced the two-fluid theory for shock acceleration
in which the momentum flux due to accelerated particles is self-consistently
included into the dynamics of the shock wave \cite{achterberg84}.  In this contribution it is
argued that if radio galaxies are the most powerful particle accelerators in
the Universe responsible for the diffuse isotropic
gamma-ray background, and
possibly also for the ultrahigh-energy cosmic rays up to 
at least $5\times 10^{19}$~eV (the so-called Greisen-Zatsepin-Kuz\'min cutoff),
they require a particle acceleration efficiency even larger than that of
supernova remnants.

\section{Energy density of the diffuse isotropic $\gamma$-ray background}

The energy losses of relativistic particles inevitably lead to
$\gamma$-radiation by which the acceleration sites can be traced.  The
cumulative flux from all unresolved cosmic accelerators appears as an diffuse
isotropic flux in an inertial frame.  A diffuse high-latitude $\gamma$-ray
background has been measured 
with CGRO from
MeV to $\sim 100$~GeV photon energies which very likely arises from unresolved
extragalactic sources \cite{sreekumar98}.  The resolved sources already contribute $\sim 15\%$ of
the diffuse isotropic flux and show the same average spectral shape as the
diffuse background.   The background
spectrum above 30~MeV as determined with EGRET on board CGRO is given by
\begin{equation} {dN\over dE}=(7.32\pm0.34)
\times 10^{-9}\left(E\over 451~{\rm MeV}\right)^{-2.1\pm0.03}
~\rm cm^{-2}~s^{-1}~MeV^{-1}~sr^{-1}.
\end{equation} 
The spectrum continues into the MeV range.
% where is has been measured with COMPTEL on board CGRO
%\cite{kappadath98}.
Most of the background flux below 10~MeV 
can be explained by the contribution of
Supernovae Ia  from the era of galaxy formation at $z_{\rm f}=2-5$
\cite{the93}.  
The uncertainties in this thermal component of the diffuse
$\gamma$-ray background are not very large, the known element abundances
tightly constrain the total number of SNIa, and hence the total energy flux in
$\gamma$-rays, in the Universe leaving little room for other sources
contributing in the MeV range.  The uncertainty in the cosmic star formation
history due to possible dust enshrouding of the earliest galaxies 
should amount to less than a factor of three.
Integrating the above non-thermal background 
spectrum between 10~MeV and 30~GeV (the flux at higher
energies must be considered somewhat unreliable due to the strongly
decreasing aperture of EGRET)
one obtains the energy density 
\begin{equation} u_\gamma=(5.03\pm 0.64)\times 10^{-6}\ {\rm eV cm^{-3}}
\end{equation} 
or 
\begin{equation} \Omega_\gamma h^2=(2.09\pm 0.27)\times 10^{-10}
\end{equation} 
where $h=H_\circ/100~\rm km~s^{-1}~Mpc^{-1}$ parametrizes the Hubble constant.  

\section{Metagalactic gamma-ray absorption and production}

Calorimetry of the particles accelerated throughout the Universe requires 
to know whether all $\gamma$-rays emitted by a source also reach the
present-day observer.   However, $\gamma$-rays of energy $E$
can interact with low-energy
photons of energy $\epsilon$ 
from the diffuse isotropic background over cosmological distance
scales $l$ producing electron-positron pairs
$\gamma+\gamma\rightarrow e^++e^-$, if their energy exceeds the
threshold energy
\begin{equation} 
\epsilon_{\rm th}={2 (m_{\rm e}c^2)^2\over
(1-\mu)(1+z)^2E}\sim 1\left(1+z\over 4\right)^{-2}\left(E\over 30~{\rm GeV}
\right)^{-1}~{\rm eV}\ \ (\mu=0) 
\end{equation} 
where $\mu$ denotes the cosine of the scattering angle
\cite{gould66,coppi97}.
The $\gamma$-ray attenuation $e^{-\tau}$ 
due to pair production becomes important
if the mean free path
$\lambda$ becomes smaller than $l$, i.e.
if the optical depth obeys $\tau=l/\lambda\ge 1$.
For the computation of $\tau$
one first needs to know the pair production cross section 
\begin{equation}
\label{sigma}
\parbox[b]{10cm}{}
\sigma_{\gamma\gamma}={3\sigma_{\rm T}\over 16}
(1-\beta^2)\left[2\beta(\beta^2-2)+(3-\beta^4)\ln\left(1+\beta\over
1-\beta\right)\right]
\end{equation}
where $\beta=\sqrt{1-1/\gamma^2}$ with $\gamma^2=\epsilon/\epsilon_{\rm th}$,
and where $\sigma_{\rm T}$ denotes the Thomson cross section
\cite{jauch76}.
Then one needs 
the geodesic radial displacement
function 
$dl/dz={c\over H_\circ}[(1+z)\bar{E}(z)]^{-1}$  
to compute the line integral from
$z=0$ to some $z=z_\circ$. For a cosmological model with $\Omega=1$
and $\Lambda=0$ the function $\bar{E}(z)$ simplifies to $(1+z)^{3/2}$.
Hence one obtains the optical depth
\begin{eqnarray}
\label{tau}
\tau_{\gamma\gamma}(E,z_\circ)=
\int_0^{z_\circ}dz{dl\over dz}\int_{-1}^{+1}d\mu{1-\mu\over 2}
\int_{\epsilon_{\rm th}}^\infty
d\epsilon n_{\rm b}(\epsilon)(1+z)^3\sigma_{\gamma\gamma}(E,\epsilon,\mu,z)
\nonumber\\
={c\over H_\circ}\int_0^{z_\circ}dz
(1+z)^{1/2}\int_{0}^{2}dx{x\over 2}\int_{\epsilon_{\rm th}}^\infty
d\epsilon n_{\rm b}(\epsilon)\sigma_{\gamma\gamma}(E,\epsilon,x-1,z)
\end{eqnarray}
adopting a non-evolving present-day background density $n_{\rm b}$,
i.e. 
$n_{\rm b}'(z,\epsilon')d\epsilon'
=(1+z)^3n_{\rm b}(\epsilon)
d\epsilon$ where the dash indicates comoving-frame quantities. 
The simplifying assumption  that the photon density transforms 
geometrically 
corresponds to a situation in which an initial short burst of star
formation at $z_{\rm f}>z_\circ$ produced most of the diffuse 
infrared-to-ultraviolet background
radiation.  Fig.1 shows the spectrum of the low-energy diffuse background
used to solve Eq.(6) numerically.
\begin{figure}
\parbox[t]{10cm}{}
\centerline{\psfig{figure=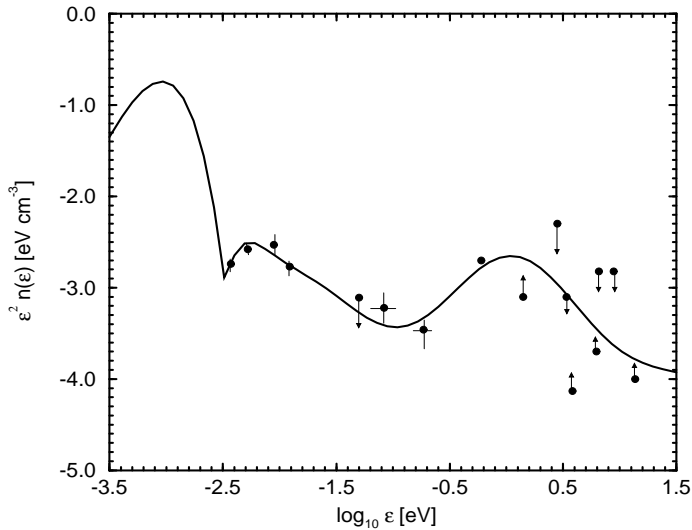,width=10cm,height=8cm}}
\caption[]{
The diffuse isotropic microwave-to-ultraviolet background: the solid
curve shows a 10th order polynomial interpolation of
observational data (\cite{fixsen98,mannheim98,franceschini98}, 
and references in \cite{madau96}).}
\end{figure}

\begin{figure}
\centerline{\psfig{figure=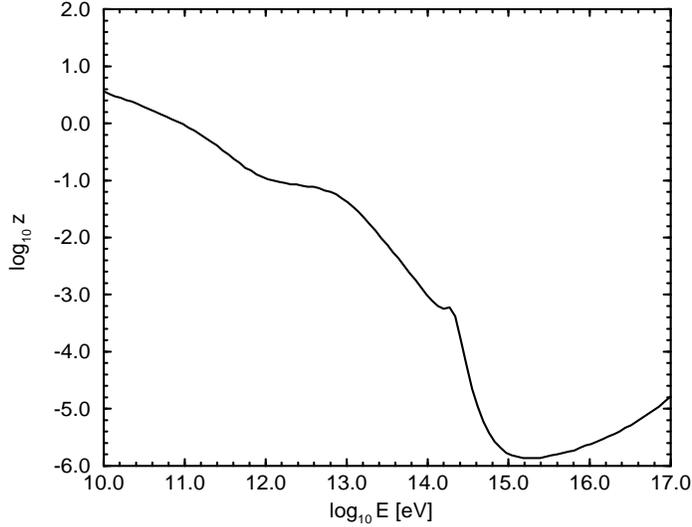,width=10cm,height=8cm}}
\caption[]{The $\gamma$-ray horizon $\tau(E,z)=1$ for the low-energy
background spectrum shown in Fig.1.  Cosmological parameters are
$h=0.6$, $\Omega=1$, and
$\Omega_\Lambda=0$. For a general discussion of pair attenuation,
see reference \cite{biller96}.}
\end{figure}
Figure 2 shows the resulting $\tau(E,z)=1$ (omitting the subscript hereafter)
curve for the microwave-to-ultraviolet diffuse background spectrum shown in
Fig.1.  It is obvious that $\gamma$-rays above $\sim 10-50$~GeV can not reach
us from beyond redshifts of $z=z_{\rm f}=2-4$.  Higher energy $\gamma$-rays
can reach us only from sources at lower redshifts (e.g.  $\gamma$-rays with
energies up to 10~TeV have been observed from Mrk~501 at $z=0.033$ in accord
with Fig.2 \cite{mannheim98}).

\vskip0.5cm
{\bf Corollary I:}  {\em If the DIGB originates from unresolved sources
distributed in redshift similar to galaxies, its spectrum must steepen above
$\sim 30$~GeV due to $\gamma$-ray pair attenuation.}
\vskip0.5cm

Here is has been tacitly assumed that the $\gamma$-rays which have turned into
electron-positron pairs do not show up again.  This is, in fact, not quite true,
since the pairs are subject to inverse-Compton scattering off the microwave
background thereby replenishing $\gamma$-rays.  The 2.7~K background is more
important as a target than the shorter wavelength background, since there is no
threshold condition for Thomson scattering 
contrary to pair production and since
2.7~K photons greatly outnumber the latter.  The inverse-Compton scattered
microwave photons turn into $\gamma$-rays of energy
\begin{equation}
E_{\rm ic}\sim 10\left(1+z\over
4\right) \left(E\over 30~{\rm GeV}\right)^2 \ {\rm MeV}
\end{equation}
conserving the energy of the absorbed $\gamma$-ray
which corresponds to a constant $E^2dN/dE$,
i.e. the expected slope of the differential spectrum
is about -2 (-2.1 observed).
A negligible amount of energy is lost to lower frequency synchrotron
emission, if magnetic fields are present in the interagalactic medium.

\vskip0.5cm
{\bf Corollary II:}  {\em Energy conservation in the reprocessing of
$\gamma$-rays from higher to lower energies by pair production and subsequent
inverse-Compton scattering produces an approximate $dN/dE\propto E^{-2}$ power
law DIGB between $\sim 10$~MeV and $\sim 30$~GeV.}
\vskip0.5cm

\section{Particle acceleration efficiency}

The
energy density of the accelerated particles in the sources ($u_{\rm acc}$) 
must obey $u_{\rm acc}\ge
u_\gamma$ for consistency.  Defining the radiative efficiency
\begin{equation}
\xi_{\rm rad}={u_\gamma\over u_{\rm acc}}={\rm Min}\left[1,{t_{\rm a}\over
t_{\rm c}}\right]
\end{equation}
where $t_{\rm a}$ and $t_{\rm c}$ denote the adiabatic and 
energy loss time scales,
respectively, the energy requirement for the sources is minimized for
$\xi_{\rm rad}=1$.  In extragalactic radio sources, $\xi_{\rm rad}\sim 1$ 
corresponds to very large electron  ($\gamma_{\rm
e}\sim 10^5$) or proton Lorentz factors ($\gamma_{\rm
p}\sim 10^{10}$) consistent with the non-thermal $\gamma$-ray
spectra extending above 10~GeV.

\begin{figure}
\parbox[t]{10cm}{}
\centerline{\psfig{figure=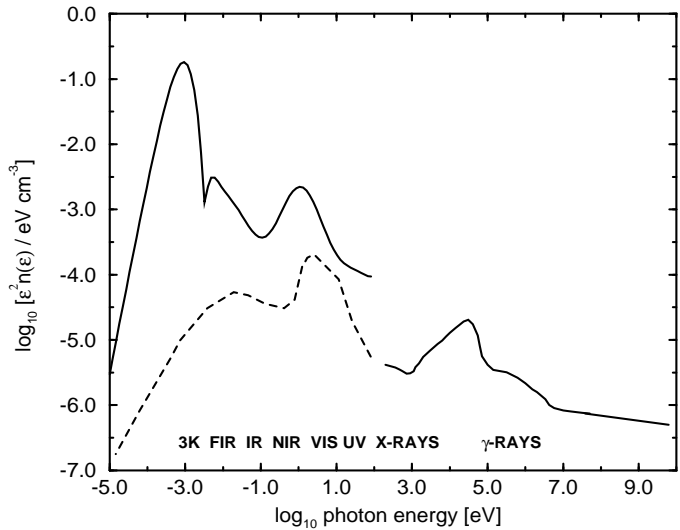,width=10cm,height=8cm}}
\caption[]{
Extragalactic background radiation.  The solid curve
interpolates observational data.  Whereas AGN dominate
in the X-ray regime, their longer wavelength emission (dotted line
from \cite{sanders89}) is 
surpassed by the stellar and dust emission from early galaxies.}
\end{figure}

The resolved extragalactic $\gamma$-ray sources belong to the so-called blazar
subclass of radio-loud active galactic nuclei
(AGN) in which one observes jets emerging from
accreting supermassive black holes preferentially at small angles to their
relativistic velocity vector.  If the entire class of radio-loud AGN is
responsible for the DIGB, the particle acceleration efficiency is constrained
by
\begin{equation}
\xi_{\rm acc}={u_{\rm acc}\over u_{\rm j}}= {u_\gamma \over \xi_{\rm rad} u_{\rm j}}
\end{equation}
where $u_{\rm j}$ denotes the total (kinetic + magnetic + randomized
relativistic particle) energy density in extragalactic jets.  Since $u_\gamma$
is known (Eq.(1)), a determination of $u_{\rm j}$ is necessary to further constrain
$\xi_{\rm acc}$.  This energy density can be inferred from the relative
strengths of the various bumps in the overall diffuse background spectrum
(Fig.3) in the following way.

\section{Origins of the diffuse isotropic background radiation}

The microwave bump is the well-known signature of the big bang
at the time of decoupling with its energy density given by
the Stefan-Boltzmann law $u_{\rm 3K}=\sigma T^4$.
The bump in the far-infrared is due to star formation in early galaxies
(at $z=z_{\rm f}$), since part of the stellar light, which is visible
as the bump at visible wavelengths, is reprocessed by
dust obscuring the star-forming regions.  The energy density of the two bumps
can be inferred from the present-day heavy element abundances.
Heavy elements have a mass fraction 
$Z=0.03$ and were produced in early bursts of star formation
by nucleosynthesis with radiative efficiency 
$\epsilon=0.007$ yielding
\begin{equation}
u_{\rm ns}\sim {\rho_* Z \epsilon c^2\over 1+z_{\rm f}}\sim
6\times 10^{-3}\left(\Omega_*h^2\over 0.01\right)\left(1+z_{\rm f}\over
4\right)^{-1}\ {\rm eV~cm^{-3}}
\end{equation}
About half of the energy is contained in either bump.
It has been shown recently by a number of
groups that probably all galaxies (except dwarfs) contain
supermassive black holes in their centers which are actively
accreting over a fraction of $t_{\rm agn}/ t_*\sim 10^{-2}$
of their lifetime implying that the electromagnetic radiation
released by the accreting black holes amounts to
\begin{equation}
u_{\rm accr}\sim {\epsilon_{\rm accr}M_{\rm bh}\over Z\epsilon M_*}
{t_{\rm agn}\over t_*}u_{\rm ns}\sim 1.4\times 10^{-4}\ \rm eV~cm^{-3}
\end{equation}
adopting the accretion efficiency $\epsilon_{\rm accr}=0.1$
and the black hole mass fraction $M_{\rm bh}/M_*=0.005$
\cite{rees98}.  Most of the accretion power emerges in
the ultraviolet where the diffuse background is unobservable
owing to photoelectric absorption by the neutral component of the
interstellar medium.  However, a fraction of $u_{\rm x}/u_{\rm bh}\sim
20\%$ (from the average quasar spectral energy distribution
\cite{sanders89}) shows up in hard X-rays due to coronal emission 
from the accretion disk to produce the diffuse isotropic X-ray background
bump with $u_{\rm x}\sim 2.8\times 10^{-5}~\rm eV~cm^{-3}$
\cite{gruber92}.

Jets with non-thermal ($\gamma$-ray) emission show up 
only in the radio-loud fraction $\xi_{\rm rl}\sim 20\%$ of
all AGN and their kinetic power roughly equals
the accretion power \cite{rawlings91}.  Hence
one obtains for the energy density in extragalactic jets
\begin{equation}
u_{\rm j}=\left(\xi_{\rm rl}\over 0.2\right)u_{\rm accr}\sim 
\left(\xi_{\rm rl}\over 0.2\right)2.8
\times 10^{-5}~\rm eV~cm^{-3}
\end{equation}
Substituting this value into Eq.(9) one obtains a limit
for the acceleration efficiency
\begin{equation}
\xi_{\rm acc}\ge 0.18\left(\xi_{\rm rad}\over 1.0\right)^{-1}
\left(\xi_{\rm rl}\over 0.2\right)^{-1}
\end{equation}
which has to be compared with the 13\% efficiency required for
supernova remnants.  Since a radiative efficiency much lower than
unity must be considered realistically (e.g., to account for the 
pdV work of the jets against a surrounding intracluster medium),
the acceleration efficiency may have to be as large
as $30\%$ or more.
Possible escapes to this conclusion are 
(i) the radio-loud fraction
was much larger at high redshifts than at low redshifts,
(ii) there
is another contributor to the gamma-ray background, such as 
the decay of topological defects \cite{sigl96}, or
(iii) the flux of the DIGB is actually lower due to
the foreground from a galactic halo \cite{dixson98}.

\section{A remark on cosmic rays from radio galaxies}

An extragalactic cosmic ray component seems to
be evident from the observed change in slope of the local spectrum
above $3\times 10^{18}$~eV.  The slope of this extragalactic component
is much steeper than $E^{-2}$ consistent with the steepening due
to energy losses for a cosmologically distributed and evolving 
source population, such as extragalactic radio sources \cite{rachen93}.
The total energy density in the loss-steepened 
extragalactic cosmic ray component and the DIGB are of the same order
of magnitude, as required if the $\gamma$-rays trace neutral pions
and the cosmic rays trace neutrons produced in photo-meson production events.

\section{Summary}

The particle acceleration efficiency of the jets in radio galaxies
must be larger than $\sim 18\%$ if the $\gamma$-rays from these
jets are to explain the DIGB.

{}

\end{document}